\documentclass[a4paper,11pt]{article}
\usepackage[margin=1in]{geometry}
\usepackage{amsmath}
\usepackage{amssymb}
\usepackage{aas_macros}
\usepackage{authblk}
\usepackage{booktabs}
\usepackage{graphicx}
\usepackage{hyperref}
\usepackage{listings}
\usepackage{multirow}
\usepackage{xcolor}

\newcommand{\TCB}{\mathrm{TCB}}
\newcommand{\TDB}{\mathrm{TDB}}

\newcommand{\TCL}{\mathrm{TCL}}
\newcommand{\M}{\mathrm{M}}

\newcommand{\LB}{L_\mathrm{B}}

\title{Lunar Time Ephemeris LTE440: User Manual}
\author[1]{Xu Lu}
\author[1]{Tian-Ning Yang}
\author[1]{Yi Xie\thanks{Email: yixie@pmo.ac.cn}}
\affil[1]{Purple Mountain Observatory, Chinese Academy of Sciences, Nanjing 210023, People's Republic of China}

\begin{document}
\maketitle

\begin{abstract}
We present the numerical lunar time ephemeris LTE440 based on the definition of Lunar Coordinate Time (TCL) given by the International Astronomical Union (IAU) in IAU 2024 Resolution II.
LTE440 can be used to obtain the numerical transformation between TCL and Solar System Barycentric Dynamical Time (TCL-TDB) or Solar System Barycentric Coordinate Time (TCL-TCB).
The theoretical model, numerical method, and performance of LTE440 are discussed.
The secular drifts between TCL and TDB and between TCL and TCB are respectively estimated as $\langle d\TCL/d\TDB\rangle=1+6.798\,355\,238\times10^{-10}$, and $\langle d\TCL/d\TCB\rangle=1-1.482\,536\,216\,67\times10^{-8}$.
The most significant periodic terms have amplitudes of about 1651 microseconds for the annual term and 126 microseconds for the monthly term.
The precision of LTE440 is estimated to be at the level of several picoseconds. LTE440 is freely available at \url{https://github.com/xlucn/LTE440}.
\end{abstract}

\section{Outline}
\label{sec:lte440}

We provide the numerical lunar time ephemeris LTE440 based on the definition of Lunar Coordinate Time (TCL) given by the International Astronomical Union (IAU) in IAU 2024 Resolution II.
LTE440 can be used to obtain the numerical transformation between TCL and Solar System Barycentric Dynamical Time (TCL-TDB) or between TCL and Solar System Barycentric Coordinate Time (TCL-TCB).
More details on Lunar Coordinate Time can be found in \cite{2024PhRvD.110h4047K,2025ApJ...985..140T}.

LTE440 consists of one BSP file \texttt{lte440.bsp} and one TPC file \texttt{lte440.tpc}.
The BSP file contains the periodic part of the lunar time ephemeris, while the coefficient of secular drift is stored in the TPC file.
The TCL-TDB transformation is given by the sum of the periodic part and the secular drift.
Detailed usage of these files is described in Section \ref{sec:usage}.

Sect. \ref{sec:build} describes the details of the construction of the lunar time ephemeris LTE440.
The theoretical model of TCL-TDB transformation is provided in Sect. \ref{sec:model}.
The model closely follows the TT-TDB model in DE440, considering the gravitational potential of the Sun, all planets, main-belt asteroids, and Kuiper belt objects.
The numerical method used is described in Sect. \ref{sec:method}.
Information on the secular drift and major periodic terms is presented in Sect. \ref{sec:performance}.

After preliminary analysis, we find that the secular drift between TCL and TDB is approximately $\langle d\TCL/d\TDB\rangle=1+6.798\,355\,238\times10^{-10}$,
and the secular drift between TCL and TCB is approximately $\langle d\TCL/d\TCB\rangle=1-1.482\,536\,216\,67\times10^{-8}$.
The most significant periodic terms are the annual term with an amplitude of about 1.65 milliseconds and the monthly term with an amplitude of about 126 microseconds.
The precision of the lunar time ephemeris is estimated to be at the level of several picoseconds.

\section{Usage}
\label{sec:usage}

\subsection{Product Files}

LTE440 consists of two files: \texttt{lte440.bsp} and \texttt{lte440.tpc}, which are freely available\footnote{The files are available at \url{https://github.com/xlucn/LTE440}}.
The published archive also includes example files \texttt{lte440.py}, \texttt{lte440.in}, and \texttt{lte440.out}.
Table \ref{tab:lte440-files} lists all the files and their descriptions.

\begin{table}[!ht]
  \caption{Descriptions of files included in the LTE440 package.}\label{tab:lte440-files}
  \centering
  \begin{tabular}{l|l|l}
    \toprule
    Filename & Content & Notes \\
    \midrule
    \texttt{lte440.bsp} & Periodic part of TCL-TDB & 1) Binary SPK kernel file \\
    & &2) Time argument is TDB\\
    & &3) Unit in seconds\\
    \texttt{lte440.tpc} & 1) Coefficient of secular drift of TCL-TDB & Text PCK kernel file \\
    &2) Body name and ID &\\
    \texttt{lte440.py} & Python example file & \\
    \texttt{lte440.in} & Input file for the example &\\
    \texttt{lte440.out} & Output generated by the example &\\
    \bottomrule
  \end{tabular}
\end{table}

When exporting the data into a SPK file, a Naif ID must be assigned.
In LTE440, TCL-TDB is assigned the ID \texttt{1000000005}
\footnote{Currently, \texttt{1000000001} is assigned for TT-TDB, \texttt{1000000002} for TCG-TCB, and \texttt{1000000003} for both TT-TDB and TCG-TCB\cite{2015jsrs.conf..254H}.
For Lunar Time Ephemeris TCL-TDB, we have assigned an ID \texttt{1000000005} for the time being.}.
The ID for the dummy center body is \texttt{1000000000}.
The coefficient of secular drift of TCL-TDB is stored in the TPC file with the name \texttt{BODY1000000005\_RATE}.

\subsection{Reading Data}
\label{sec:interface}

Data from the BSP and TPC files can be accessed using the SPICE toolkit\footnote{https://naif.jpl.nasa.gov/naif/toolkit.html} or CALCEPH\footnote{https://calceph.imcce.fr}.
The example files illustrate how to read the periodic terms and secular drift data, and calculate the numerical values of TCL-TDB.
In the Python file \texttt{lte440.py}, the functions \texttt{tclmtdb()} and \texttt{tclmtcb()} are provided to calculate TCL-TDB and TCL-TCB, respectively.
The relevant descriptions are shown in Table \ref{tab:functions}.

\begin{table}[!ht]
  \caption{Description of the functions in the example Python file.}\label{tab:functions}
  \centering
  \begin{tabular}{l|l|l|l|l}
    \toprule
    \multirow{2}{*}{Function} & \multicolumn{2}{c|}{Input} & \multicolumn{2}{c}{Output} \\
    \cmidrule{2-5}
    & Definition & Unit & Definition & Unit \\
    \midrule
    \texttt{tclmtdb()} & TDB epoch & Julian Date & TCL-TDB & second \\
    \texttt{tclmtcb()} & TDB epoch & Julian Date & TCL-TCB & second \\
    \bottomrule
  \end{tabular}
\end{table}

The example code below shows a simple usage of these functions.

\begin{lstlisting}[
  language=Python,
  caption={Example usage of LTE440 in Python}]
import lte440
import spiceypy as spice

jd_tdb = 2451545.0  # J2000.0 TDB
spice.furnsh(['lte440.bsp', 'lte440.tpc'])

tcl_tdb = lte440.tclmtdb(jd_tdb)
print(f"TCL-TDB={tcl_tdb:+.16e}")
# TCL-TDB=+4.9330749643254812e-01

tcl_tcb = lte440.tclmtcb(jd_tdb)
print(f"TCL-TCB={tcl_tcb:+.16e}")
# TCL-TCB=-1.0760479771816941e+01

jd_tcl = jd_tdb + tcl_tdb / 86400
print(f"TCL={jd_tcl:+.16e}")
# TCL=+2.4515450000057095e+06
\end{lstlisting}

\section{Model and Numerical Method}
\label{sec:build}

\subsection{Model of TCL-TDB transformation}
\label{sec:model}

According to the definition of Lunar Coordinate Time (TCL) in IAU 2024 Resolution II\footnote{IAU 2024 Resolution II: "to establish a standard Lunar Celestial Reference System (LCRS) and Lunar Coordinate Time (TCL)", \url{https://www.iau.org/Iau/Publications/List-of-Resolutions}}, the transformation between TCL and TDB is given by the TCG-TDB relation, with the Earth-related quantities replaced by the Moon-related quantities. With the transformation between TDB and TCB that \footnote{IAU 2006 Resolution B3: "Re-definition of Barycentric Dynamical Time, TDB", \url{https://www.iau.org/Iau/Publications/List-of-Resolutions}}
\begin{equation}
	\TDB=\TCB-\LB(\TCB-T_0)+\TDB_0,
  \label{eq:tcbmtdb}
\end{equation}
the TCL-TDB relation can be obtained as
\begin{eqnarray}
  \TDB-\TCL&=&-\frac{\LB}{1-\LB}(\TDB-T_0)+\frac{1}{1-\LB}\TDB_0 \nonumber \\
  &&+\frac{1}{1-\LB}\left[\frac{1}{c^2}\int_{\TDB_0+T_0}^\TDB\left(\frac{v_\M^2}{2}+w_\mathrm{0M}+w_{l\M}\right)\mathrm{d}t+\frac{1}{c^2}\boldsymbol{v}_\M\cdot(\boldsymbol{x}-\boldsymbol{x}_{\M})\right]\nonumber \\
  &&-\frac{1}{1-\LB}\left[\frac{1}{c^4}\int_{\TDB_0+T_0}^\TDB\left(-\frac{v_\M^4}8-\frac{3}{2}v_\M^2w_\mathrm{0M}+4\boldsymbol{v}_\M\cdot\boldsymbol{w}_{\M}+\frac12w_\mathrm{0M}^2+\Delta_\M\right)\mathrm{d}t\right.\nonumber\\
  &&-\frac{1}{c^4}\left.\left(3w_\mathrm{0M}+\frac{v_\M^2}2\right)\boldsymbol{v}_\M\cdot(\boldsymbol{x}-\boldsymbol{x}_{\M})\right].
  \label{eq:tdbmtcl}
\end{eqnarray}
The time argument in Eq. \eqref{eq:tdbmtcl} is TDB, and all variables on its right side are TDB-compatible.
$w_\mathrm{0M}$ is the gravitational potential of external bodies at the Moon's center, given by
\begin{equation}
  w_\mathrm{0M}=\sum\limits_\mathrm{A\neq M}\frac{GM_\mathrm{A}}{r_\mathrm{MA}}.
\end{equation}
with the summation over all bodies A other than the Moon. The bodies considered here include the Sun, all planets, 343 main-belt asteroids, 30 Kuiper belt objects (KBO), and the KBO ring representing the rest of the Kuiper belt, which are the same as those considered in DE440\cite{Park2021AJ161.105}.
$w_{l\M}$ is the potential due to external oblate figures
\begin{equation}
  w_{l\M}=-\frac{GM_\mathrm{S}J_\mathrm{2S}R^2_\mathrm{S}}{2r^3_\mathrm{MS}}(3\sin{\phi^2_\mathrm{M,S}}-1)-\frac{GM_\mathrm{E}J_\mathrm{2E}R^2_\mathrm{E}}{2r^3_\mathrm{ME}}(3\sin{\phi^2_\mathrm{M,E}}-1).
\end{equation}
Here only the contribution from the Sun and the Earth are included. $\phi_\mathrm{M,S}$ and $\phi_\mathrm{M,E}$ are the latitudes of the Moon relative to the Sun's equator and the Earth's equator, respectively. $\boldsymbol{w}_{\M}$ and $\Delta_\M$ are defined as
\begin{equation}
  \boldsymbol{w}_{\M}=\sum\limits_\mathrm{A\neq M}\frac{GM_\mathrm{A}\boldsymbol{v}_\mathrm{A}}{r_\mathrm{MA}},
\end{equation}
and
\begin{equation}
  \Delta_\M=\sum\limits_\mathrm{A\neq M}\frac{GM_\mathrm{A}}{r_\mathrm{MA}}\left\{\sum_{\mathrm{B}\neq \mathrm{A}}\frac{GM_\mathrm{B}}{r_{\mathrm{BA}}}-2v_\mathrm{A}^2+\frac12\left[\frac{\left(\boldsymbol{r}_\mathrm{MA}\cdot\boldsymbol{v}_\mathrm{A}\right)^2}{r_\mathrm{MA}^2}+\boldsymbol{r}_\mathrm{MA}\cdot\boldsymbol{a}_\mathrm{A}\right]\right\}.
\end{equation}
with the summation over the Sun and all planets.

\subsection{Numerical Method}
\label{sec:method}

We have numerically evaluated the time integral part of TCL-TDB given by Eq. \eqref{eq:tdbmtcl}.
The orbital data are obtained from the JPL DE440/DE441 ephemerides\cite{Park2021AJ161.105}.
The numerical integration is performed using the 10th-order Romberg integration algorithm, with a step size of 0.5 days.
We have fitted the result with 13th-degree Chebyshev polynomials described in \cite{Newhall1989CelMech45.305} with 4-day granules.
To validate the precision of the numerical method, it is applied to calculate TT-TDB in DE440 and the results agree to within 1 picosecond.

To preserve maximum numerical precision, we fitted the secular drift and obtained the periodic part by subtracting the secular term.
The periodic part is stored in a BSP file and the coefficient of secular drift is stored in a TPC file. Thus, the calculation of TCL-TDB from LTE440 requires an extra addition of the secular term in comparison to TT-TDB from DE440.

\subsection{Performance}
\label{sec:performance}

Using a preliminary least-squares fit, we estimate the secular drift of TCL-TCB to be approximately
\begin{equation}
  \left\langle\frac{d\TCL}{d\TCB}\right\rangle=1-1.482\,536\,216\,67\times10^{-8}\pm1.0\times10^{-17},
  \label{eq:lcm}
\end{equation}
and the secular drift of TCL-TDB as
\begin{equation}
  \left\langle\frac{d\TCL}{d\TDB}\right\rangle=1+6.798\,355\,238\times10^{-10}\pm1.0\times10^{-17},
  \label{eq:lbm}
\end{equation}
where the errors are the least-squares fitting errors.
The comparison of the results with those from \cite{2024PhRvD.110h4047K} and \cite{2025ApJ...985..140T} is shown in Table \ref{tab:tclmtdb-rate}.

\begin{table}[!ht]
  \caption{Comparison of secular drift of TCL-TCB}\label{tab:tclmtdb-rate}
  \centering
  \begin{tabular}{l|l|l}
    \toprule
    $\left\langle\frac{d\TCL}{d\TCB}\right\rangle$ & Work & Notes \\
    \midrule
    $1-1.482\,536\,216\,67\times10^{-8}$ & Our work & \\
    $1-1.482\,4\times10^{-8}$ & \cite{2024PhRvD.110h4047K} & Calculated from the value $1.280\,8\,$ms/day therein \\
    $1-1.482\,536\,24\times10^{-8}$ & \cite{2025ApJ...985..140T} & \\
    \bottomrule
  \end{tabular}
\end{table}

By applying fast Fourier transform (FFT), we find that there are 13 periodic terms in TCL-TDB with amplitudes above 1 microsecond.
The amplitudes $A_i$, periods $T_i$, and phases $\phi_i$ of these terms $A_i\sin(2\pi{T_i}^{-1}(t-\mathrm{J}2000.0)+\phi_i)$ are shown in Table \ref{tab:lunarte-short}.
Note that the errors of the results are limited by the frequency resolution of FFT.

\begin{table}[!ht]
  \caption{Information on major periodic terms in TCL-TDB}
  \label{tab:lunarte-short}
  \centering
  \begin{tabular}[c]{r|r|r|r}
    \toprule
    Index  $i$ & Amplitude $A_i$ [$\mu$s] & Period $T_i$ [day] & Phase $\phi_i$ [rad] \\
    \midrule
    1 & 1651.36355077 & 365.26590909 & 3.10895165 \\
    2 & 126.30813184 & 29.53053800 & 5.18472464 \\
    3 & 19.37467715 & 398.99950348 & 1.33855843 \\
    4 & 13.70088760 & 182.63295455 & 3.07602294 \\
    5 & 7.47520418 & 411.67264344 & 3.32446352 \\
    6 & 4.24397312 & 4320.34946237 & 3.43186281 \\
    7 & 3.76051430 & 377.97977422 & 0.92358639 \\
    8 & 2.93368121 & 14.25402654 & 1.09317212 \\
    9 & 2.67752983 & 369.63431463 & 1.51225314 \\
    10 & 2.36687890 & 32.12797857 & 5.21748801 \\
    11 & 1.85820098 & 10859.25675676 & 2.56843762 \\
    12 & 1.09742615 & 584.00072674 & 4.67635157 \\
    13 & 1.08850698 & 292.00036337 & 2.99248981 \\
    \bottomrule
  \end{tabular}
\end{table}

\section*{Acknowledgments}
This work is funded by the National Natural Science Foundation of China (Grants No. 62394350, No. 62394351 and No. 12273116), the Strategic Priority Research Program on Space Science of the Chinese Academy of Sciences (XDA300103000, XDA30040000, XDA30030000).

\bibliography{refs}

\begin{thebibliography}{1}

\bibitem{2024PhRvD.110h4047K}
Sergei~M. {Kopeikin} and George~H. {Kaplan}.
\newblock {Lunar time in general relativity}.
\newblock {\em \prd}, 110(8):084047, October 2024.

\bibitem{2025ApJ...985..140T}
Slava~G. {Turyshev}, James~G. {Williams}, Dale~H. {Boggs}, and Ryan~S. {Park}.
\newblock {Relativistic Time Transformations between the Solar System
  Barycenter, Earth, and Moon}.
\newblock {\em \apj}, 985(1):140, May 2025.

\bibitem{2015jsrs.conf..254H}
J.~{Hilton}, C.~{Acton}, J.~E. {Arlot}, S.~{Bell}, N.~{Capitaine}, A.~{Fienga},
  W.~{Folkner}, M.~{Gastineau}, D.~{Pavlov}, E.~{Pitjeva}, V.~{Skripnichenko},
  and P.~{Wallace}.
\newblock {Report of the IAU Commission Working Group on Standardizing Access
  to Ephemerides and File Format Specification: Update September 2014}.
\newblock In Z.~{Malkin} and N.~{Capitaine}, editors, {\em Journ{\'e}es 2014
  ``Syst{\`e}mes de r{\'e}f{\'e}rence spatio-temporels''}, pages 254--255,
  August 2015.

\bibitem{Park2021AJ161.105}
Ryan~S. {Park}, William~M. {Folkner}, James~G. {Williams}, and Dale~H. {Boggs}.
\newblock {The JPL Planetary and Lunar Ephemerides DE440 and DE441}.
\newblock {\em \aj}, 161(3):105, March 2021.

\bibitem{Newhall1989CelMech45.305}
X.~X. {Newhall}.
\newblock {Numerical Representation of Planetary Ephemerides}.
\newblock {\em Celestial Mechanics}, 45:305, January 1989.

\end{thebibliography}
\bibliographystyle{unsrt}
\end{document}